\documentclass[runningheads]{llncs}
\usepackage[T1]{fontenc}
\usepackage{graphicx}
\usepackage{color}

\usepackage{cite}
\usepackage{amsmath,amssymb,amsfonts}
\usepackage{algorithmic}
\usepackage{textcomp}
\usepackage{xcolor}
\usepackage{multirow}

\begin{document}
\title{Learning Heuristics for the Maximum Clique Enumeration Problem Using Low Dimensional Representations\thanks{Supported by the T\"UB\.ITAK Project 118E283.}}
\titlerunning{Learning Heuristics for the Maximum Clique Enumeration Problem}
%
\author{Ali Baran Taşdemir\inst{1,2}\orcidID{0000-0001-5271-5751} \and
Tuna  Karacan\inst{1,3,4}\orcidID{0000-0003-3591-7057} \and
Emir Kaan K\i rmac\i \inst{1,3,4}\orcidID{0000-0002-5352-0363}\and 
Lale \"Ozkahya\inst{1,3}\orcidID{0000-0001-6105-1694}}

\authorrunning{Taşdemir et al.}
%
\institute{Department of Computer Engineering, Hacettepe University
Ankara, Turkey\\
\and \email{alibaran@tasdemir.us} \\
\and \email{\{tunakx,ekaankirmaci,lale.ozkahya\}@gmail.com}\\
\and Equal contribution
}
\maketitle              

\begin{abstract}
Approximate solutions to various NP-hard combinatorial optimization problems have been found by learned heuristics using complex learning models. In particular, vertex (node) classification in graphs has been a helpful method towards finding the decision boundary to distinguish vertices in an optimal set from the rest. 
By following this approach, we use a learning framework for a pruning process of the input graph 
towards reducing the runtime of the maximum clique enumeration problem. 
We extensively study the role of using different vertex representations on the performance of this heuristic method, using graph embedding algorithms, such as Node2vec and  DeepWalk, and representations using higher-order graph features comprising local subgraph counts. 
Our results show that 
Node2Vec and DeepWalk are promising embedding methods in representing nodes 
towards classification purposes. 
We observe that using local graph features in the classification process produce more accurate results when combined with a feature elimination process. 
Finally, we provide tests on random graphs  to show the robustness and scalability of our method. 
\keywords{Maximum clique enumeration \and Node classification \and Node embedding.}
\end{abstract}

\section{Introduction}\label{sec:intro}
Graphs are natural objects to represent complex data in real-life relations, such as social networks, molecular networks, and finance networks. Most challenging problems arise in solving combinatorial optimization questions in a time-efficient way, since most of them are known to be NP-hard. The maximum clique enumeration problem (MCE) seeks to enumerate all subgraphs of maximum size, where all vertices are neighbors of each other.  The MCE problem is NP-hard and it is a strengthening of the maximum clique problem, which 
aims to find the size of the maximum clique only. 
The maximum clique problem is a well-studied problem and known to be NP-complete among other strong hardness results~\cite{chen2006strong,zuckerman2006linear}. 
As many NP-complete problems defined on graphs arise 
as real-life problems on networks, the MCE problem is also 
applicable in the analysis of social network~\cite{wasserman1995social}, 
behavioral networks \cite{bernard1979informant}, 
financial networks \cite{boginski2005statistical}, and 
citation and dynamic networks \cite{stix2004finding}.

Recently, there has been studies such as \cite{dutta2019finding, grassia2019learning} finding estimate solutions for these problems using a two-stage framework: 1) embedding the graph vertices into low-dimensional vectors; 2) using machine learning to reduce the size of the input graph via vertex classification for finding an approximate solution efficiently. This approach is shown to be applicable for very dense and 
large real-world networks and scalable to different domains. 


\paragraph{Our contribution:} 
In this work, we build upon this approach by studying the role of using different vertex representations towards estimating the solution of the MCE problem. 
We conduct our experiments two-fold, representing vertices using: 1) graph embedding algorithms: Node2vec \cite{grover2016node2vec}, 
DeepWalk \cite{perozzi2014deepwalk}, and GraphSage \cite{hamilton2017inductive}, 2) local frequencies of higher-order graph structures as learning features. 
We make use of these methods above to represent 
the vertices in the input graph as low-dimensional vectors. 
Later these vectors are used in a binary classification 
of the vertices to predict which vertices are in a maximum clique. 
This helps to remove the vertices that are least likely 
to be in a maximum clique from the graph so that the input 
size is reduced. 

Our results show that 
Node2Vec and DeepWalk are promising embedding methods in representing nodes 
towards classification purposes. We also observe that using local counts of 
graphs in Fig. \ref{fig:all_orbits} as vertex features provides high classification 
accuracy when combined with a feature elimination process. 
Our method is tested on real world networks to show the high accuracy of 
our results. We show the robustness and scalability of our results on random graphs.


\section{Framework}\label{sec:methodology}

\subsection{Low dimensional Vertex Representations}
\begin{figure}[h]
    \centering
    \includegraphics[scale=0.045]{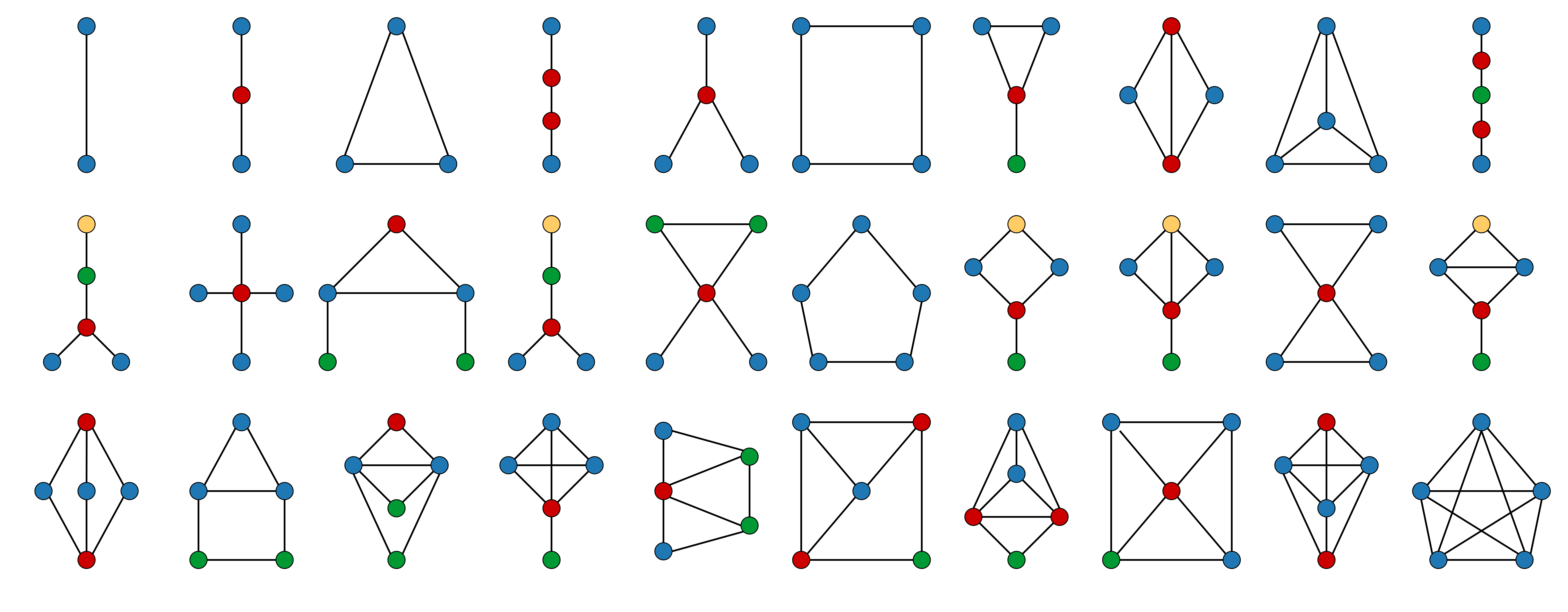}    
    \caption{The 5-vertex subgraphs and orbits whose local counts are used as  graph features using the Evoke algorithm in~\cite{pashanasangi2020efficiently}.}
    \label{fig:all_orbits}
\end{figure}

As the initial step, we obtain the vector representation of each vertex 
in the input graph. 
Graph embeddings are used to transform the vertices of a graph to vectors and 
a good embedding is expected to capture graph topology, such as degree 
distribution, and clustering coefficient. Thus, we use 3 different embedding 
methods, Node2vec \cite{grover2016node2vec}, DeepWalk \cite{perozzi2014deepwalk}, and GraphSage \cite{hamilton2017inductive}, to represent the vertices as vectors. 
The first two use a strategy based on random walks preserving the connectivity information between a vertex and its neighbors. 
GraphSage is an algorithm focusing on inductive learning tasks and represents vertices by aggregating features from their neighbors.
In addition to embedding techniques, we use the local frequencies of all 5-vertex induced subgraphs (graphlets) in Fig. \ref{fig:all_orbits} as vertex features, 
using the state-of-the-art algorithm Evoke~\cite{pashanasangi2020efficiently}. 
As seen in Fig. \ref{fig:all_orbits}, mostly there is more than one role/position of a vertex in a graphlet as distinguished by different colors in Fig. \ref{fig:all_orbits}. Each of these positions is called an {\it orbit} and each vertex occurs in one of the orbits in a graphlet. The {\it orbit frequency} at a vertex is the number of times a vertex occurs in one of these orbits and each of these frequencies is considered as a distinct vertex feature. 

\subsection{Classification Procedure}
The vertices are classified according to their presence in a maximum clique. If the vertex is in a maximum clique, we label that vertex with label-1. If the vertex is not in a maximum clique, we label that vertex with label-0. The purpose of this classification is to find the vertices that will be pruned from the graph. 
The goal is to train a model by obtaining a mapping $\gamma:V\to \{0,1\}$  using the training set as $T=\{<f(v_i),y_i>\}_{i=1}^L$ with $L$ samples,  
where $f(v_i)$ indicates the feature vector associated with 
$v_i\in V$ and $y_i\in\{0,1\}$ is the class label for for that vertex. 

In this learning process, a probabilistic classifier $P$ is used 
to obtain a probability distribution over $\{0,1\}$ for every 
given $f(u),$ $u\in V.$ 
With the help of this and defining a {\it confidence threshold} $q\in [0,1]$, 
the vertex set is pruned to reduce the size of the graph as follows. 
By choosing a well-performing confidence threshold $q$, 
the vertices to prune are defined by the set 
$V'=\{u\in V: P(u=1)\leq q\}.$ In that sense, 
picking a higher value of $q$ implies a higher pruning ratio. 
We experiment with various classifiers via 
scikit-learn that is an open-sourced library providing tools for supervised and unsupervised machine learning~\cite{pedregosa2011scikit}. 
To obtain the confidence values, we use predict-proba method in the scikit-learn implementation which outputs the probability that 
a vertex is in a maximum clique.

{\it Vertex Classification:} 
For each network $G_{i}$ in the training set, we list all maximum cliques $\mathcal{C}_i = \{C_1,C_2,...,C_{n}\}$ in $G_{i}$,  and label the vertices contained in these cliques with 1. To create a balanced dataset, we use the in-sample approach and randomly pick 1.5 times of this many vertices from  $G_i\setminus \mathcal{C}_i$ and label them with 0. We use mainly two classifier algorithms based on their performance in the experiments: logistic regression and random forest. 
For logistic regression, we train with a 5-fold cross-validation for a maximum of 100 epochs,
using one of L1 or L2 regularizers and 0.5 or 1 as the regularization term multiplier determined by a grid search.
For random forest, we train with a 5-fold cross-validation with 5, 25, or 100 estimators and using "gini" or "entropy" split quality function determined by a grid search.

{\it Evaluation Metrics:}  
The {\it pruning ratio} $P$ describes the overall ratio of the number of vertices 
predicted to not be in a solution with probability at least $q$. 
The {\it clique number} of a graph $G$ is the number of vertices in a maximum clique of $G$, denoted by $\omega (G).$ 
The {\it clique accuracy} is 
the overall ratio of the number of graphs $G$ with $\omega (G)=\omega (G')$ in a sampling set, where 
$G'$ is the graph obtained by pruning. 

\section{Experimental Results}

In the experiments, we use the state-of-the-art algorithm cliquer \cite{niskanen2003cliquer} to find cliques, which uses the branch-and-bound algorithm of \cite{ostergaard2002fast}. To calculate clique counts on graphs, we use a computer with Amd Epyc 7B12 (2.25 GHz) processor and 102 GB memory, running Debian 9. Cliquer requires large memory for big networks. Hence, alternatively, we also use the igraph-python library \cite{csardi2006igraph} 
to enumerate maximum cliques, which is a software implementation of a modified Bron-Kerbosch algorithm \cite{eppstein2010listing}. For that process, we use a computer with an Intel Core i7 processor and 32 GB of memory.

\subsection{Results on Real-World Graphs}\label{sec:real}

For testing the performance of our framework on real-world networks, we chose three network domains from~\cite{nr}: biological networks, social networks, and web networks, that contain 26, 93, and 12 networks, respectively. 
We train a distinct classifier for each 
particular representation method and particular network domain. 
The classifiers for the biological networks, social networks, and web networks use a sample of 25, 90, and 10 networks, respectively. The F1 scores for all  classifiers are above 80\% (around 90\% in average over this group) except classifiers trained using Graphsage embedding vectors and Evoke vertex feature vectors on the Social networks domain, for which the F1 score is around 55\%. 

\begin{table}[t]
\caption{The clique number ($\omega$), pruning ratio ($P$), runtime in seconds by igraph ($T_i$) results on real-world networks (with $|V|$-$|E|$) using $q=0.4$. Representation methods as Node2Vec (N2V), Deepwalk (DW), GraphSage (GS), and graph features obtained via Evoke (Ev.), respectively. }
\label{tab:realworld-table}
\centering
\begin{tabular}{l|c|c|c|c|c|c|c|c|c}
\multicolumn{1}{c}{} &
\multicolumn{3}{c}{bio-WormNet-v3} & 
\multicolumn{3}{c}{web-wikipedia2009} &  
\multicolumn{3}{c}{web-google-dir}\\
\multicolumn{1}{c}{} &
\multicolumn{3}{c}{(16K - 763K)} &
\multicolumn{3}{c}{(2M - 5M)} &  
\multicolumn{3}{c}{(876K - 5M)}\\ 
\multicolumn{10}{c}{} \\
&\textbf{$\omega$} & $P$ & \textbf{$T_i$}  
&\textbf{$\omega$} & $P$ & \textbf{$T_i$} 
&\textbf{$\omega$} & $P$ & \textbf{$T_i$} \\ \hline
orig. 
& 121 & - & 7044 	
& 31 & - & 312 
& 44 & - & 63	\\ 
N2V 
& {\bf 111}   & {\bf 0.77} 	& 5324 	
& 25   	& 0.92 	& 267   
& {\bf 34}   	& {\bf 0.85} 	& 54   \\
DW 
& 78  	& 0.6 	& 0.2
& {\bf 30}    & {\bf 0.88} 	& 275   	
& 30    & 0.88 	& 52   	\\ 
GS 
& 104 	& 0.40 	& 448
& 19   	& 0.83 	& 262  
& 29   	& 0.81 	& 52 	\\
Ev.
& 61  & 0.94 	& 3.2
& 9   & 0.97 	& 160    
& 20  & 0.98 	& 35 \\ 
\cite{lauri2019fine} 	
& 90 	& 0.90 	& -		
& 31 	& 0.99 	& -			
& 44 	& 0.97 	& -	 \\  

\multicolumn{10}{c}{}

\end{tabular}

\begin{tabular}{l|c|c|c|c|c|c|c|c|c}
\multicolumn{1}{c}{} & 
\multicolumn{3}{c}{socfb-A-anon} & 
\multicolumn{3}{c}{socfb-B-anon} & 
\multicolumn{3}{c}{socfb-Texas84}\\
\multicolumn{1}{c}{} & 
\multicolumn{3}{c}{(3M - 24M)} & 
\multicolumn{3}{c}{(3M - 21M)} &  
\multicolumn{3}{c}{(36K - 2M)}\\
\multicolumn{10}{c}{}  \\
& \textbf{$\omega$} & $P$ & \textbf{$T_i$}  
& \textbf{$\omega$} & $P$ & \textbf{$T_i$} 
& \textbf{$\omega$} & $P$ & \textbf{$T_i$} \\ \hline
orig. 				
& 25	& -		& 1317 
& 24	& -		& 1166 	
& 51	& -		& 98 	\\    
N2V  				
& 11    & 0.99 	& 1287 	   
& 18    & 0.99 	& 921  
& {\bf 49}   	& {\bf 0.93} 	& 23    \\
DW  				
& {\bf 18}    & {\bf 0.94} 	& 992  
& {\bf 23}    & {\bf 0.75} 	& 935     
& 42   	& 0.38 	& 3.2   \\ 
GS 				
& 18    & 0.73 	& 1081   
& 16    & 0.74 	& 865   
& 27   	& 0.72 	& 0.9  \\
Ev.     				
& 16  & 0.69 	& 698        
& 17  & 0.71 	& 683     
& 29  & 0.77 	& 2.3     \\  
\cite{lauri2019fine} 	
& 23 	& 0.94 	& -			
& 23 	& 0.94 	& -			
& 44 	& 0.97 	& -	\\ 

\multicolumn{10}{c}{}
\end{tabular}%
\end{table}

{\it Node Embedding Specifications:} 
We use the python implementations of Node2vec and DeepWalk 
in \cite{n2v} and \cite{openne}. 
For both DeepWalk and Node2vec, the representation sizes of all vertices, the walk number, the walk length, the skip-gram window size, and the number of parallel processes are set to 128, 5, 10, 10 and 8, respectively. 
The value of the return parameter $p$ is chosen as 0.25 and the in-out parameter $q$ as 0.75. 
For GraphSage, we perform unsupervised learning by using the mean aggregator. We set the representation size for embedding to 50 and use two layers with the size of 50 and L2 normalization after each layer. We use randomly initiated features with a length of 32. 
The parameters listed above are not optimized and fine tuning the values  would help to improve the accuracy and runtime.


We present the results on real-world networks in Table \ref{tab:realworld-table}, where 
the computation cost of enumerating all maximum cliques in the pruned graph is listed as $T_i$. 
The optimal confidence threshold is decided to be $q=0.4$ considering the tradeoff between the accuracy and pruning ratio. 
We present the performance of four different representation methods and compare it with the state-of-the-art method used in \cite{lauri2019fine} together with the exact values listed as 'original'. 
For each test network, the best performing model using our method is highlighted. We observe that classifiers using graph embeddings by Node2Vec and DeepWalk as feature vector perform better compared to the other models used for our method. Our method differs from the state-of-the-art method in \cite{lauri2019fine} also by not having a preprocessing stage, which uses a degree-method to eliminate the vertices with relatively low degrees. In that sense, Node2Vec and DeepWalk are observed as promising embedding methods in representing nodes towards classification purposes. 

We also observe that the use of higher-order node features needs 
fine tuning of its parameters such as the subset of features used.  
In the following section, we introduce a feature elimination method that 
greatly improves the performance of the classifiers when using the local node 
features introduced above. 


\subsection{Feature Elimination Procedure}



In order to improve the performance of the results obtained by orbit counts as features, we conduct a feature elimination process. 
We apply a combination of different feature selection steps, which are RFE (Random Feature Elimination), 
Univariate Selection, and Pearson Correlation, 
using the scikit-learn library. 
Higher-order graph features are selected by using a combination of these techniques. Features are ranked by using each method from 1 to $n$ in descending order, $n$ being the total number of features. 
The scores given by these different techniques are averaged to obtain a combined score for each feature. 
Initially, we experiment with feature vectors without 
ranking them by their performance in each network domain, called 
 "f1", "f2", and "f3". 
We determine three threshold scores $s_1$, $s_2$ and $s_3$ such that $s_1>s_2>s_3$.  
 Then, $fi$ ($i\in \{1,2,3\}$) contains all vectors with a score above $s_i$. 
 Hence, we have $f1 \subset f2 \subset f3 \subset f0$, where being a subset means containment relation in terms of the set of features in the corresponding vector and $f0$ is the vector of all features. 
Next, we redo these evaluations obtaining scores when the performance 
of the features are restricted to a particular domain.   
We introduce the domain-specific vectors as $fx1$, $fx2$, and $fx3$, 
where $x$ takes the value $b$, $w$, and $s$ for the biology, web, and social networks domain, respectively.

\begin{table*}[!htbp]
\centering
\caption{The information loss vs. pruning ratio for the corresponding feature vectors. The information loss is given by the error percentage in the predicted clique number.}
\label{tab:FE-realworld}
\resizebox{\textwidth}{!}{%
\begin{tabular}{l|l|l|l|l|l|l|l|l|l|l|l|l|l|l}
\multicolumn{1}{c|}{\multirow{2}{*}{Network}} &
  \multicolumn{14}{c}{} \\
\multicolumn{1}{c|}{} &
  \multicolumn{2}{c|}{f0} &
  \multicolumn{2}{c|}{f1} &
  \multicolumn{2}{c|}{f2} &
  \multicolumn{2}{c|}{f3} &
  \multicolumn{2}{c|}{fx1} &
  \multicolumn{2}{c|}{fx2} &
  \multicolumn{2}{c}{fx3} \\ \hline
bio-WormNet-v3 &
  0.50 &
  \multicolumn{1}{l|}{0.94} &
  \textbf{0.00} &
  \multicolumn{1}{l|}{\textbf{0.80}} &
  {\bf 0.00} &
  \multicolumn{1}{l|}{{\bf 0.81}} &
  \textbf{0.00} &
  \multicolumn{1}{l|}{\textbf{0.81}} &
  0.00 &
  \multicolumn{1}{l|}{0.75} &
  \textbf{0.00} &
  \multicolumn{1}{l|}{\textbf{0.81}} &
  \textbf{0.00} &
  \textbf{0.81} \\
web-wikipedia-2009 &
  0.71 &
  \multicolumn{1}{l|}{0.97} &
  0.68 &
  \multicolumn{1}{l|}{0.96} &
  0.45 &
  \multicolumn{1}{l|}{0.87} &
  \textbf{0.35} &
  \multicolumn{1}{l|}{\textbf{0.97}} &
  \textbf{0.39} &
  \multicolumn{1}{l|}{\textbf{0.97}} &
  \textbf{0.35} &
  \multicolumn{1}{l|}{\textbf{0.95}} &
  0.68 &
  0.98 \\
web-google-dir &
  0.55 &                        
  \multicolumn{1}{l|}{0.98} &
  \textbf{0.27} &
  \multicolumn{1}{l|}{\textbf{0.94}} &
  0.52 &
  \multicolumn{1}{l|}{0.96} &
  0.52 &
  \multicolumn{1}{l|}{0.97} &
  \textbf{0.27} &
  \multicolumn{1}{l|}{\textbf{0.97}} &
  0.55 &
  \multicolumn{1}{l|}{0.97} &
  0.52 &
  0.97 \\
socfb-A-anon &
  0.36 &
  \multicolumn{1}{l|}{0.69} &
  0.20 &
  \multicolumn{1}{l|}{0.72} &
  \textbf{0.12} &
  \multicolumn{1}{l|}{\textbf{0.73}} &
  0.12 &
  \multicolumn{1}{l|}{0.71} &
  0.16 &
  \multicolumn{1}{l|}{0.73} &
  0.16 &
  \multicolumn{1}{l|}{0.58} &
  \textbf{0.08} &
  \textbf{0.71} \\
socfb-B-anon &
  0.29 &
  \multicolumn{1}{l|}{0.71} &
  0.17 &
  \multicolumn{1}{l|}{0.71} &
  \textbf{0.17} &
  \multicolumn{1}{l|}{\textbf{0.72}} &
  0.13 &
  \multicolumn{1}{l|}{0.69} &
  \textbf{0.17} &
  \multicolumn{1}{l|}{\textbf{0.71}} &
  0.17 &
  \multicolumn{1}{l|}{0.62} &
  0.21 &
  0.60 \\
socfb-Texas84 &
  0.43 &
  \multicolumn{1}{l|}{0.77} &
  \textbf{0.27} &
  \multicolumn{1}{l|}{\textbf{0.75}} &
  0.29 &
  \multicolumn{1}{l|}{0.75} &
  0.27 &
  \multicolumn{1}{l|}{0.73} &
  0.51 &
  \multicolumn{1}{l|}{0.81} &
  0.20 &
  \multicolumn{1}{l|}{0.63} &
  0.31 &
  0.62\\
\end{tabular}%
}
\end{table*}

In Table \ref{tab:FE-realworld}, we observe that feature elimination procedure increases 
the accuracy in the predicted clique number when the values are compared with those under $f0$. 
In this table, we highlight the cases that produce slightly better results than the reference case $f0$ by reducing the information loss remarkably while having comparable pruning ratios.  
Moreover, the domain-independent feature vectors $f1$, $f2$ and $f3$ 
seem to provide results that are at least as accurate as the domain-specific 
feature vectors.

\subsection{Robustness and Scalability}\label{sec:synthetic}

The experiments presented below aim to show that our method is able to accurately enumerate maximum cliques even in the cases when the training is done on small instances of graphs or when there is only one maximum clique in the input graph. For that, we generate random graphs for training and testing our model by planting a maximum clique of a particular size in each of them. 

The random graph $G(n,p)$, also known as Erd{\H o}s-Renyi graph~\cite{paul1959random} 
is a graph on $n$ vertices, where every edge exists with probability $p.$ 
We sample all random graphs using the Erd{\H o}s-Renyi model with $G(n,1/2)$. To plant a clique of size $k$, we sample $k$ vertices uniformly at random  and insert edges between all vertex pairs in this set. Thus, we guarantee that each network in the dataset contains at least one clique with size at least $k$.

For every $n$, the value of the expected clique number $k$ is $2\log(n)$. Hence, in the training stage, we sample $G(n,1/2)$ with a planted clique of the order close to $k$ and generate 300 different graphs for the training set corresponding to each different pair $(n,k)$. We use the same process for each of the three embedding methods.

{\it Vertex Classification:} 
The vertices are represented by the feature vectors comprising the local frequencies of all orbits listed in Fig. \ref{fig:all_orbits}.  
In the training process, we use all vertices with label-1 (in max-clique) and sample 1.5 times of this amount from vertices with label-0 (not in max-clique). 
Although efficient algorithms exist
~\cite{feldman2017statistical} 
for finding 
these cliques, 
it is still an open problem to efficiently find the cliques whose order are in the range between 
$\sqrt{n}$ and $2\log(n).$ Hence, to show that our method is also able to produce accurate and efficient results in that range, we pick values in our experiments as 
$(n,k)\in \{(128, 12), (256, 13), (512, 15)\}$, using sample graphs with the same value of $n$ and $k$. 

\begin{table}[!htbp]
\centering
\caption{Robustness with fixed $n$ and increasing $k$, using random forest as the classifier algorithm and $q =0.4$. 
For each planted clique of order $k+1$, $k+2$ and $k+3$, the pruning ratio, $P$, and the clique accuracy, $A_C$, show average values.}
\label{tab:randomtests_q40}
\begin{tabular}{cccccccc}
\hline
\multicolumn{2}{c}{Training Sets} 
& \multicolumn{2}{c}{k + 1}                                          
& \multicolumn{2}{c}{k + 2}                                          
& \multicolumn{2}{c}{k + 3}                     \\ \hline
n     & \multicolumn{1}{c|}{k}    & $P$ 	& $A_C$ 	& $P$ 	& $A_C$ 	 		& $P$ 	& $A_C$  	\\ \hline
128   & \multicolumn{1}{c|}{12}   & 0.94 	& 0.73    		& 0.93 		& 0.96   		& 0.93 		& 1.0   	\\
256   & \multicolumn{1}{c|}{13}   & 0.89 	& 0.34 				& 0.89 		& 0.48   				& 0.89 		& 0.54   		\\
512   & \multicolumn{1}{c|}{15}   & 0.87 	& 0.05 			& 0.87 		& 0.15 			& 0.87 		& 0.20 	\\ \hline
\end{tabular}%
\end{table}
In the testing stage, we use random graphs with a 
growing planted clique size as $k'=k+1, k+2, k+3,$ 
where for each value of $k'$ 100 samples are used, shown in Table \ref{tab:randomtests_q40}. 
The clique accuracy indicates the ratio of the sample graphs whose maximum clique size does not change after the pruning stage. 
As expected, the performance of the classifiers improves for larger values of $k'.$ This is also 
observed on the performance of the algorithms solving 
the planted clique problem for large $k$~\cite{jerrum1992large,kuvcera1995expected}. 
For larger values of $n$ the accuracy drops, since 
existence of a planted clique causes 
a more significant deviation from the 
expected value of $\omega$ as $n$ increases. 
We observe speedups of the order 100x when $q=.40$ is used.

\section{Conclusions} 




We propose a heuristic method to find an approximate solution of the Maximum Clique Enumeration problem by narrowing the search space. Our experiments show that graph embedding algorithms Node2Vec and DeepWalk are suitable choices for representing the vertices in this framework as well as using higher-order graph features. 
Our method is also shown to be robust and scalable by extensive experiments on real world and random networks. 
The accuracy and runtime results can be further improved by optimizing the hyperparameters in the embedding process and incorporating preprocessing techniques. 

\subsubsection{Acknowledgements} 
This research was supported in part 
by the T\"UB\.ITAK Project 118E283. 

%
%
%
\bibliographystyle{splncs04}
\bibliography{SubmissionCSoNet2022}

\end{document}